\documentclass[%
reprint,
superscriptaddress,
amsmath,
amssymb,
aip,
floatfix,
jap]{revtex4-2}
\usepackage{graphicx}
\usepackage{dcolumn}
\usepackage{bm}
\usepackage[utf8]{inputenc}
\usepackage[T1]{fontenc}
\usepackage{mathptmx}
\usepackage{textcomp}
\usepackage{siunitx}
\usepackage{bigints}

\usepackage[switch]{lineno}
\usepackage[table,xcdraw,dvipsnames]{xcolor}
\usepackage{booktabs}
\usepackage{float}
\usepackage{color}
\usepackage{ulem}
\usepackage{multirow}
\begin{document}

\title[Curvature-stabilized three-dimensional spiral states and local-field nucleation of skyrmions on spherical caps]{Curvature-stabilized three-dimensional spiral states and local-field nucleation of skyrmions on spherical caps}

\author{Sabri~Koraltan}
\email{sabri.koraltan@tum.de}
\affiliation{Zentrum f\"ur QuantumEngineering, Technische Universit\"at M\"unchen, D-85748 Garching, Germany}
\affiliation{Institute of Applied Physics, TU Wien, Vienna, A-1040, Austria}
\affiliation{Physics of Functional Materials, Faculty of Physics, University of Vienna, A-1090 Vienna, Austria}%
\affiliation{Institute of Physics, University of Augsburg, D-86135 Augsburg, Germany}
\author{Joe~Sunny}
\affiliation{Institute of Physics, University of Augsburg, D-86135 Augsburg, Germany}
\author{Emily~Darwin}
\affiliation{Magnetic \& Functional Thin Films Laboratory, Empa, Swiss Federal Laboratories for Materials Science and Technology, Ueberlandstrasse 129, 8600 Dübendorf, Switzerland}
\author{Daniel~Rothhardt}
\affiliation{Magnetic \& Functional Thin Films Laboratory, Empa, Swiss Federal Laboratories for Materials Science and Technology, Ueberlandstrasse 129, 8600 Dübendorf, Switzerland}
\author{Reshma~Peremadathil-Pradeep}
\affiliation{Magnetic \& Functional Thin Films Laboratory, Empa, Swiss Federal Laboratories for Materials Science and Technology, Ueberlandstrasse 129, 8600 Dübendorf, Switzerland}
\author{Micha{\l}~Krupi{\'n}ski}
\affiliation{Institute of Nuclear Physics Polish Academy of Sciences, Radzikowskiego 153, Krakow 31-342, Poland}
\author{Takeaki~Gokita}
\affiliation{Institute of Applied Physics, TU Wien, Vienna, A-1040, Austria}
\author{Robert~Kraft}
\affiliation{Physics of Functional Materials, Faculty of Physics, University of Vienna, A-1090 Vienna, Austria}%
\affiliation{Vienna Doctoral School in Physics, University of Vienna, A-1090 Vienna, Austria}%
\author{Claas~Abert}
\affiliation{Physics of Functional Materials, Faculty of Physics, University of Vienna, A-1090 Vienna, Austria}%
\author{Jakub~Jurczyk}
\affiliation{Institute of Applied Physics, TU Wien, Vienna, A-1040, Austria}
\author{Amalio~Fernández-Pacheco}
\affiliation{Institute of Applied Physics, TU Wien, Vienna, A-1040, Austria}
\author{Markus~Weigand}
\affiliation{Institut für Nanospektroskopie, Helmholtz–Zentrum Berlin für Materialien und Energie GmbH, 12489 Berlin, Germany}
\author{Sebastian~Wintz}
\affiliation{Institut für Nanospektroskopie, Helmholtz–Zentrum Berlin für Materialien und Energie GmbH, 12489 Berlin, Germany}
\author{Dieter~Suess}
\affiliation{Physics of Functional Materials, Faculty of Physics, University of Vienna, A-1090 Vienna, Austria}%
\author{Hans~Josef~Hug}
\affiliation{Magnetic \& Functional Thin Films Laboratory, Empa, Swiss Federal Laboratories for Materials Science and Technology, Ueberlandstrasse 129, 8600 Dübendorf, Switzerland}
\author{Manfred~Albrecht}
\affiliation{Institute of Physics, University of Augsburg, D-86135 Augsburg, Germany}

\date{\today}

\begin{abstract}
The combination of polystyrene particles and multilayers with chiral interfacial interactions offers an interesting platform to study the interplay between topology, chirality, and curvature. Here, we investigate Pt/Co/Ta multilayers sputtered onto self-assembled polystyrene particles with a diameter of $\SI{919}{nm}$, which show a magnetic reversal distinctly different from that of planar films. Using non-invasive scanning transmission X-ray microscopy and in-vacuum magnetic force microscopy (MFM) with low-moment tips, we observe maze domains in the planar films, whereas the curvilinear films host three-dimensional spiraling stripe states. Repeated scanning with a conventional higher-moment MFM tip ruptures these spirals into metastable skyrmions at the top of the particles. Micromagnetic simulations of spherical caps reveal that increasing curvature renders a single spiral stripe energetically favorable over an isolated skyrmion, and that this stabilization is predominantly of magnetostatic origin. Our results identify the spiral states as intrinsic to the curvilinear geometry and the skyrmions as metastable states nucleated by local magnetic fields, underlining the value of non-invasive imaging to distinguish intrinsic curvature effects from probe-induced magnetic textures.
\end{abstract}

\maketitle

\section{Introduction}
Magnetic skyrmions are a central focus in condensed matter physics~\cite{back20202020, koraltan2026skyrmionics}, due to their topological properties~\cite{nagaosa2013topological}, and potential applications~\cite{finocchio2016magnetic} in memory~\cite{fert2013skyrmions}, logic~\cite{finocchio2015skyrmion}, sensing~\cite{al2016skyrmion, lianeris2025spintronic, koraltan2024skyrmionic}, or unconventional computing devices~\cite{lee2024task, lee2023perspective, dacamarasantaclaragomes2025neuromorphic}. In recent years, many different stabilization mechanisms and material systems have been investigated~\cite{jiang2015blowing, montoya2017tailoring, heigl2021dipolar, hassan2024dipolar, Okubo_PhysRevLett.108.017206, kurumaji2019skyrmion}. Originally, crystals with chiral interactions and the bulk Dzyaloshinskii–Moriya interaction (DMI) were utilized to study Bloch skyrmions at low temperatures and high magnetic fields~\cite{muhlbauer2009skyrmion, yu2010real}. Another common way to nucleate skyrmions is to harness interfacial DMI in heavy metal/ferromagnetic bilayers and multilayers~\cite{jiang2015blowing, boulle2016room, moreau-luchaire2016additive, soumyanarayanan2017tunable}. On the basis of the chosen materials, thicknesses, or repetition numbers, it is possible to tune the existence of skyrmions and the conditions for the formation of skyrmion lattices~\cite{jefremovas2025role}. 

Theoretical studies show that curvilinear surfaces can give rise to a curvature-induced chirality, typically referred to as curvature-induced DMI~\cite{carvalho2015stability, kravchuk2018multiplet, yang2021intrinsic, yang2021robust}. Theoretical works suggest that for sufficiently large curvatures, spherical nanoshells can lead to the formation of magnetic skyrmions, depending on the interplay between interfacial DMI, curvature, and perpendicular magnetic anisotropy (PMA)~\cite{yang2021intrinsic}. However, it is important to note that the exchange-driven curvature-induced DMI scales as $D_{\rm curv} \propto A/R$, where $A$ is the exchange stiffness and $R$ is the radius of curvature~\cite{gaididei2014curvature, sheka2015curvature}. For experimentally accessible particle sizes, with radii on the order of hundreds of nanometers, $D_{\rm curv}$ is expected to be very small compared to typical interfacial DMI values. In addition to these local exchange-driven terms, curving the geometry also modifies the magnetostatic interaction, giving rise to nonlocal chiral effects, whose relative importance increases with the film thickness~\cite{sheka2020nonlocal} and which are therefore expected to be particularly relevant for thick multilayers.

Over the past two decades, self-assembled polystyrene (PS) particles have provided a reliable platform to study spherical nanocaps~\cite{streubel2016magnetism}, where early realizations dealt with the use of Co/Pd multilayers to study curvature-induced modifications on the magnetic domain morphology~\cite{albrecht2005magnetic} and the formation of magnetic vortices using soft magnetic materials~\cite{streubel2012magnetic, aravind2019bistability}. Recently, skyrmionic states were observed on spherical particles coated with symmetric Pt/Co/Pt multilayers, in which the interfacial DMI of the top and bottom interfaces cancels in the planar case, and it was proposed that the curvature breaks this symmetry and thereby induces a sizable interfacial DMI~\cite{dugato2025curved}. Furthermore, it was shown that PS particles can be utilized to geometrically constrain the number of skyrmions in Ir/Fe/Co/Pt skyrmion hosting multilayers, where the number of skyrmions can be controlled by the size of the particles~\cite{sam2025magnetic}. However, in these studies, MFM was the main imaging technique, complemented by electron holography in Ref.~\cite{dugato2025curved}. Since the local magnetic fields of MFM tips have been shown to nucleate skyrmions even in planar films~\cite{casiraghi2019individual, zhang2018magnetic}, non-invasive imaging methods are important to disentangle intrinsic curvature effects from textures induced by the measurement technique itself.

In this work, we use magnetic thin film multilayers grown on self-assembled PS particles to investigate the nucleation of skyrmions. For this purpose, we sputter-deposited a Pt/Co/Ta multilayer, which is known to generate magnetic maze domains in the demagnetized state~\cite{he2018evolution, wang2019construction}. After a comparison of the magnetic hysteresis between the planar and curvilinear films, we directly imaged the magnetic domain configuration of planar films by scanning transmission X-ray microscopy (STXM) and of curvilinear films by high-sensitivity in-vacuum magnetic force microscopy (MFM) under applied magnetic fields. Using the local magnetic field from a higher magnetic moment tip under tapping mode in a conventional MFM, the evolution and stabilization of spin textures is investigated. Using non-invasive imaging, we first show that the curvilinear geometry gives rise to three-dimensional spiral states, which are distinct from the maze domains observed in planar films. Despite the exchange-driven curvature-induced DMI being expected to be very small for the particle diameters used here, these 3D spiral states serve as precursors from which skyrmions can be nucleated under repeated scanning with the MFM tip, decaying into isolated skyrmions at the top of the spheres. To understand the energetics behind these observations, we complement our experiments with micromagnetic simulations of spherical caps whose curvature is varied continuously from a nearly flat disk to a strongly curved cap, similar to previous theoretical studies of curvilinear shells~\cite{yershov2015controllable}. We find that increasing curvature renders a single spiral stripe energetically more favorable than an isolated skyrmion, and that this stabilization is predominantly of magnetostatic origin.

This paper is organized as follows. In Sec.~\ref{sec:exp}, we describe the experimental methods used to fabricate and characterize the samples, as well as the micromagnetic model. The experimental and numerical results are presented in Sec.~\ref{sec:results} and discussed in Sec.~\ref{sec:discussion}, before we conclude in Sec.~\ref{sec:conclusion}.

\section{Methods\label{sec:exp}}
\subsection{Nanosphere Lithography}
Highly ordered monolayers of submicrometer PS particles were fabricated using self-assembly at the water-air interface~\cite{li2021recent}. At the beginning of this process, the utilized substrate was treated in a soft isotropic nitrogen-oxygen plasma for 5 min to remove contaminations and enhance wettability. Shortly after plasma treatment, the substrate was submerged under a surface of distilled water inside a Petri dish, where the self-assembly process was carried out. The monodisperse aqueous suspension of PS particles with an average diameter $d$ of 919 nm (standard deviation of 26 nm) from MicroParticles GmbH Berlin was diluted by mixing with an equal volume of ethanol (99.5\% pure). The suspension was applied to the surface of the water using a curved glass Pasteur pipette with a rate of approximately $\SI{1}{\micro\litre\per\second}$. When the entire surface of the water in the Petri dish was filled with PS particles, 2D crystallization was forced by rocking the Petri dish. This resulted in the formation of standing waves on the water surface leading to a highly ordered close-packed monolayer of the particles. The monolayers were transferred to the substrate by slow evaporation of water at room temperature. The completion of the process was confirmed by scanning electron microscopy (SEM)~\cite{sam2024size, aravind2019bistability, sam2025magnetic}.

\subsection{Thin Film Growth}
Magnetic multilayers with the following stack, sub./${\rm [Pt(3\,nm)/Co(2\,nm)/Ta(2\,nm)]_{12}}$, were deposited at room temperature from elemental targets on both planar Si/SiOx substrates and Si/SiOx substrates covered with PS particles with $d = \SI{919}{nm}$. Furthermore, the planar system was also grown on $\SI{30}{nm}$ thick $\rm{Si_3N_4}$-membranes (9 square windows with an edge size of $\SI{100}{\micro\metre}$), suitable for X-ray-based imaging techniques, as we will discuss later. The base pressure of the BESTEC sputtering system was better than $\SI{5e-8}{mbar}$. We used Ar as a sputter gas at a working pressure of $\SI{3.5}{\micro\bar}$. The deposition rates for each material were determined by a calibrated quartz microbalance system. The samples were rotated continuously during the sputtering process to ensure homogeneous film deposition. In the remainder of the paper, we will refer to thin films grown on flat Si/SiOx substrates and $\rm{Si_3N_4}$-membranes as \textit{planar}, and to films grown on PS particles as \textit{spheres}.

\begin{figure*}[htbp]
    \centering
    \includegraphics[width=0.95\linewidth]{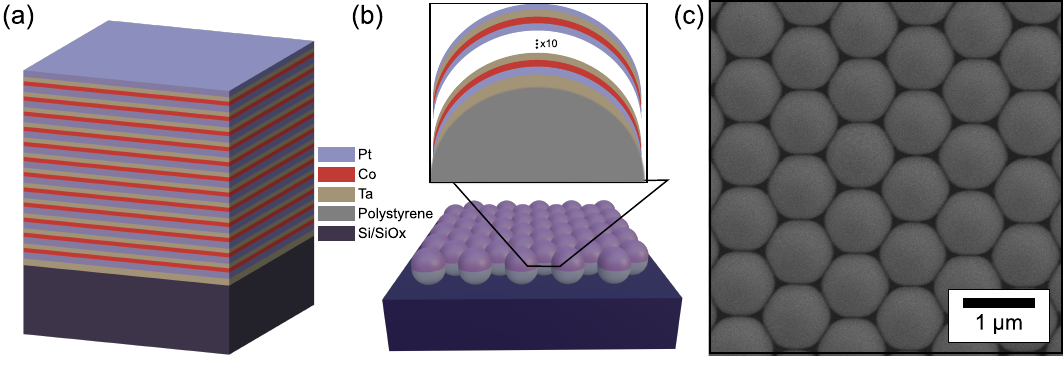}
    \caption{(a) Illustration of the multilayer stack sub./${\rm [Pt/Co/Ta]_{12}}$. (b) Schematic of the multilayer stack sputtered onto the particle monolayer with a zoomed-in cross-section, which highlights the existence of a radial thickness gradient across the surface of the spherical particle. (c) SEM image of the particle assembly (averaged particle diameter: $\SI{919}{nm}$).}
    \label{fig:fig1}
\end{figure*}

\subsection{Vibrating Sample Magnetometry~\label{subsec:vsm}}
Magnetic characterization was performed using a LakeShore vibrating sample magnetometer (VSM) at room temperature. The sample was mounted on a measurement rod, and a built-in rotation motor was used to change the direction of the magnetic field applied onto the magnetic specimen by effectively changing the angle between the sample's plane and the magnetic field generated by a water-cooled electromagnet.

\subsection{Scanning Transmission X-ray microscopy~\label{subsec:stxm}}
The planar films prepared on the $\rm{Si_3N_4}$-membranes were imaged in transmission non-invasively by STXM at the MAXYMUS endstation~\cite{weigand2022timemaxyne} at the BESSY II electron storage ring operated by the Helmholtz-Zentrum Berlin für Materialien und Energie. The images are acquired at the Co $\rm L_3$-edge corresponding to an energy of $\SI{780.2}{eV}$, where the energy was slightly optimized for maximum contrast. The direction of the X-rays is perpendicular to the plane of the sample. Thus, the X-ray magnetic circular dichroism (XMCD) signal can be used to obtain a contrast proportional to the out-of-plane (oop) component of the magnetization. Typically, XMCD can be obtained as the difference between two images taken using X-rays with positive and negative circular polarization. However, the XMCD contrast is already available in each of the images taken with one polarization, revealing the two oop magnetization orientations. To reduce measurement time and isolate the magnetic contrast from non-magnetic absorption, we used an image of the saturated state as a reference signal, which was subtracted from each image.

\subsection{Magnetic force microscopy~\label{subsec:mfm}}
Two different MFM setups were used for different purposes. In the following, we detail their usage, main differences, and purposes.

\paragraph{High-sensitivity vacuum MFM:} Here, we rely on a home-built high-sensitivity MFM operated under vacuum conditions~\cite{Feng2022, mandru2020coexistence, koraltan2025signatures}. To optimize the signal-to-noise ratio and spatial resolution, we utilize high-aspect ratio tip Nanotec SS-ISC uncoated cantilevers with a nominal tip radius below $\SI{5}{nm}$. The resonance frequency of the cantilever was measured to be $f_{c} = \SI{55}{kHz}$. The magnetic sensitivity of the tip was introduced by depositing a Ta(2~nm)/Co(4.5~nm)/Ta(3~nm) trilayer on the tip side facing the cantilever chip~\cite{Feng2022}. The deposition was carried out using a shadow mask to prevent the coating near the cantilever base, thereby preserving a high quality factor of $Q \approx 1.5 \times 10^{5}$. This provides an improved measurement sensitivity that allows MFM with low magnetic moment tips~\cite{Feng2022}. The cantilever was excited at resonance using a phase-locked loop, with an oscillation amplitude $A = 5$~nm. The detection channel corresponds to the resonance frequency shift $\Delta f$, while the phase remained stable at $-90^{\circ}$. In this configuration, attractive magnetic interactions, such as those generated between an up-magnetized tip and the stray field of an up-oriented domain, manifest as negative frequency shifts, which are represented as blue contrast in the raw data. Measurements were performed in single-pass mode, with tip-sample spacing stabilized by keeping the amplitude of the second side bands constant at $f_c \pm 2f_{ac}$, which is proportional to $\dfrac{\rm{d^2}C}{\rm{d}z^2}$. Thus, this method is used as a non-invasive method, which allows us to accurately map the stray fields by following a curvilinear geometry without modifying the underlying magnetic structure.

\paragraph{Conventional MFM:} Here, we used a commercially available Dimension Icon MFM from Bruker, where the cantilever oscillation is driven at a fixed frequency, and the tip-sample interaction generates a change of the phase. The measurement is performed using the Lift Mode, which employs a double-pass tapping mode performed at small distance (50~nm), while the tip intermittently contacts the sample surface to map the topography. 

An MFM measurement performed under ambient conditions leads to a sensitivity limited by the low quality factor of the cantilever. Thus, tips with stronger magnetic moment must be employed to obtain a sufficient signal-to-noise ratio. We use the conventional MFM tips from Bruker (MESP-V2). Moreover, during the topography scan, the tip contacts the surface intermittently. Hence, the strong magnetic field of the tip likely modifies the micromagnetic state of the sample. Alternatively, the strong local field of the tip can be used to facilitate the formation of vortices, magnetic domains, stripes, and even skyrmions~\cite{gartside2018realization, garanin2018writing, casiraghi2019individual}.

\subsection{Micromagnetic simulations\label{subsec:sim}}
To better understand the energetics of the spiral and skyrmion states on the curved surfaces, we performed micromagnetic simulations with a finite-element code, which is based on magnum.fe~\cite{abert2013magnumfe}. We discretize the magnetic film into tetrahedral elements with a characteristic size of $\SI{4}{nm}$, which is below the exchange length of the material. We model the curvilinear film as a spherical cap of constant thickness $t = \SI{84}{nm}$, corresponding to the total thickness of the $\rm [Pt/Co/Ta]_{12}$ stack, with a fixed geodesic radius $s = \SI{721.8}{nm}$ measured along the surface from the top of the cap to the rim. The cap is wrapped onto a sphere of radius $R = s/(\pi\kappa)$, such that its opening half-angle is $\theta_{\rm max} = \pi\kappa$. Thus, $\kappa \rightarrow 0$ describes a flat disk, and the curvature of the cap increases with $\kappa$. Note that we neglect the radial thickness gradient of the deposited film, as illustrated in Fig.~\ref{fig:fig1}(b), in order to obtain a clear isolation of the curvature effects. In this work, we consider $\kappa = 0.01$, $0.1$, $0.2$, $0.3$, and $0.4$ (see Fig.~\ref{fig:fig5}(a)). Here, $\kappa = 0.4$ is the most realistic case for our samples, since the particles are arranged in a close-packed lattice, such that the deposited nanocap does not extend all the way down to the equator of the particles.

The multilayer is treated as an effective medium~\cite{woo2016observation, lemesh2017accurate}, in which the magnetic parameters of the Co layers are scaled by the volume fraction $f = t_{\rm Co}/(t_{\rm Pt} + t_{\rm Co} + t_{\rm Ta}) = 2/7$. Note that the effective medium model might not be perfectly suitable for increasing curvature, since the individual layers of the stack experience different radii of curvature. For the Co layers, we use a saturation magnetization $M_s = \SI{1300}{kA/m}$, an exchange stiffness constant $A = \SI{15}{pJ/m}$, and an interfacial DMI constant $D = \SI{0.75}{mJ/m^2}$, which are typical values for Co in Pt-based multilayers, resulting in the effective values $fM_s$, $fA$, and $fD$. Within the effective-medium approach, the exchange and interfacial DMI energies of the Co layers are distributed over the total thickness of the stack, which is why $A$ and $D$ are scaled by the same volume fraction as $M_s$~\cite{woo2016observation, lemesh2017accurate}. The uniaxial anisotropy constant is determined from the in-plane saturation field of the planar film, $\mu_0 H_K \approx \SI{430}{mT}$, as $K_u = f M_s \mu_0 H_K/2 + \mu_0 (f M_s)^2/2$. Both the anisotropy axis and the interfacial DMI are oriented along the local surface normal $\hat{\bm{n}}$ of the cap. We include energy contributions from exchange, demagnetization, uniaxial anisotropy, interfacial DMI, and an external magnetic field $\mu_0 H_z$ applied along the $z$-axis into the effective field. Since the magnetization is computed in the full three-dimensional geometry of the curved film, both types of curvature effects are included automatically: the local curvature-induced DMI and anisotropy are contained in the exchange energy, whereas the nonlocal curvature-induced effects are contained in the demagnetization energy, which is obtained from the full magnetostatic interaction of the curved film~\cite{sheka2020nonlocal}.

We start our computations from three initial states: a spiral stripe domain that winds twice around the top of the cap, a N\'eel skyrmion with a radius of $\SI{60}{nm}$ located at the top, and a uniform state with the magnetization pointing along the local surface normal. We relax the magnetization with respect to its total energy using an appropriate energy minimization routine until the maximum torque falls below $\SI{1}{kA/m}$.

\begin{figure}[htbp]
    \centering
    \includegraphics[width=0.95\linewidth]{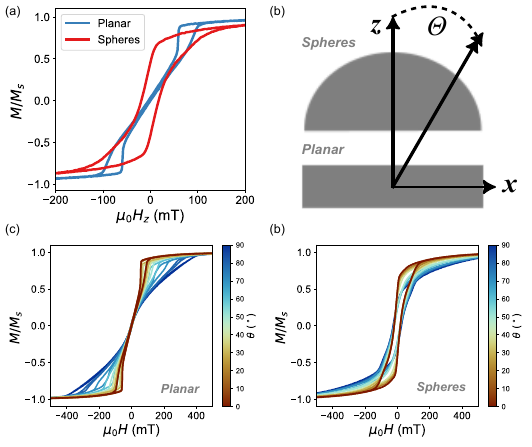}
    \caption{Comparison of magnetic properties of planar and curvilinear films, where we show the M-H hysteresis loop as a function of applied oop field $\mu_0H_z$ in (a). The measurement geometry is explained in (b), for the loops presented in (c, planar), and (d, spheres), where we measured the loops for different angles of applied magnetic fields.}
    
    \label{fig:fig2}
\end{figure}

\section{Results\label{sec:results}}
\subsection{Magnetic multilayers on spheres}
The stack used for this study is shown schematically in Fig.~\ref{fig:fig1}(a). We chose Pt/Co/Ta multilayers that exhibit PMA and typically show a magnetic maze domain pattern in the remanent state. This film system is also known to host skyrmions~\cite{wang2019construction}. The illustration of the spheres covered with the magnetic film in Fig.~\ref{fig:fig1}(b) shows that the spheres are arranged in a hexagonal lattice and are covered only in the top half, leading to curvilinear spherical nanocaps as the effective geometry studied in this work. The zoomed-in cross-sectional drawing presented in Fig.~\ref{fig:fig1}(b) schematically depicts the expected radial thickness gradient across the top surface of the sphere~\cite{sam2025magnetic} and highlights the curved profile of the thin film. As the film thickness gradually decreases from the top towards the rim of the particles, the intermixing is expected to form a soft magnetic Pt–Co–Ta alloy at the rim~\cite{sam2025magnetic,streubel2016magnetism}. The SEM image of Fig.~\ref{fig:fig1}(c) displays a monolayer of particles with magnetic coating acquired after the film deposition.

\subsection{Magnetic Characterization}
To characterize the magnetic response of the planar samples and spheres, we performed VSM measurements at room temperature. The normalized magnetization measured ($M / M_s$) as a function of applied oop magnetic field ($\mu_0H_z$) for both the planar sample (blue) and the spheres (red) is shown in Fig.~\ref{fig:fig2}(a). The M-H hysteresis loop of the planar sample behaves as expected for such a multilayer~\cite{wang2019construction}: When the magnetic field is lowered from positive saturation, the magnetization drops significantly and rather abruptly due to the formation of magnetic domains. Further lowering the field suggests that the magnetic domains change their size linearly in this range until full saturation. Furthermore, we observe the same behavior when approaching from negative saturation. This leads to the well-known butterfly-shaped hysteresis loops with two hysteretic openings~\cite{wang2019construction}.

The M-H hysteresis loop of the spheres is substantially different from that of the planar sample. The nonhysteretic part vanishes and a large hysteresis opening is observed. Remanence is not zero as for the planar case, but also not $1$ as often obtained by a system with PMA, where the magnetization state jumps from one saturated state to the other. Instead, the magnetization drops about $25\%$ at remanence and continues to decrease (increase) when the magnetic field is further lowered (increased). A similar behavior was observed for Ir/Fe/Co/Pt multilayers deposited on PS particles with diameters lower than $\SI{300}{nm}$ by Sam et al.~\cite{sam2025magnetic}.

To gain more insight about the magnetic properties of the samples, we measured the M-H loops as a function of the relative angle between the sample normal and the applied magnetic field, as illustrated in Fig.~\ref{fig:fig2}(b). The angle $\theta = \SI{0}{^\circ}$ corresponds to an oop magnetic field perpendicular to the sample surface and the angle $\theta = \pm\SI{90}{^\circ}$ describes a magnetic field applied along the $\pm x$-direction of the sample.

The M-H loops for the planar sample for angles between $\SI{0}{^\circ} \leq \theta \leq  \SI{+90}{^\circ}$ are shown in Fig.~\ref{fig:fig2}(c). First, the saturation field is increasing significantly with increasing $\theta$. The saturation field for $\theta = \SI{0}{^\circ}$ is $\mu_0H_s \approx \SI{100}{mT}$. In the case of $\theta = \SI{90}{^\circ}$, the saturation field increases to $\mu_0H_s \approx \SI{430}{mT}$. The increased in-plane saturation field is a result of the anisotropy field of the films, and it is expected to be in this range compared to the literature~\cite{sucksmith1954magnetic, dieny2017perpendicular}. 

This picture changes notably for the spheres. The M-H loops presented in Fig.~\ref{fig:fig2}(d) show a more linear and gradual increase of the saturation fields. This can be attributed to the more isotropic distribution of the perpendicular magnetic anisotropy, which is expected to remain normal to the curvilinear surface. Furthermore, the magnetic hysteresis diminishes, while a more pronounced increase in coercive fields can be observed, compared to planar counterparts.

\begin{figure*}[]
    \centering
    \includegraphics[width=\linewidth]{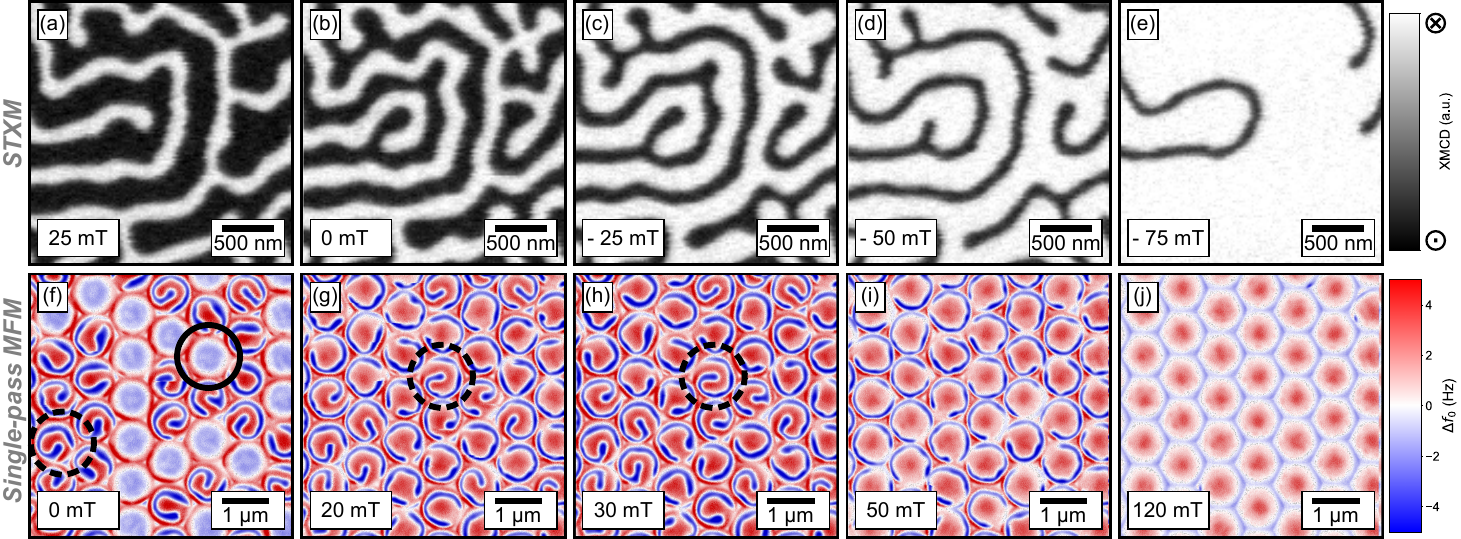}
    \caption{(a-e) Magnetic domain configuration on planar films under different oop magnetic fields imaged by STXM at the Co $L_3$ edge, where the black contrast shows domains that are magnetized up, and white contrast indicates domains that are magnetized down. (f-j) The magnetic states on the spheres are imaged using a single-pass MFM, where the distance between the tip and the surface is kept constant, thus, following the topography. The colors indicate the frequency shift, $\Delta f_0$, which is a measure of the local stray field gradients.}
    \label{fig:fig3}
\end{figure*}

\subsection{Magnetic imaging\label{subsec:res_qmfm}}
In Fig.~\ref{fig:fig3}(a-e) the STXM images of a planar sample are displayed under different oop magnetic fields after positive saturation. At $\SI{+25}{mT}$ (Fig.~\ref{fig:fig3}(a)), we have a predominantly dark contrast with several white stripes. At remanence (Fig.~\ref{fig:fig3}(b)) the size and number of the black and white domains are rather similar. Further decreasing the magnetic field shows that the magnetization is slowly reversed by shrinking the domains that are magnetized up (dark contrast), eventually leading to a saturated state for field magnitudes above $\SI{100}{mT}$. This behavior is in good agreement with our expectation from the M-H loops presented in Fig.~\ref{fig:fig2}(a).

To image the magnetic states on the spherical particles, we used the high-sensitivity in-vacuum MFM. The data displayed in Fig.~\ref{fig:fig3}(f--j) were acquired with an up tip after negative saturation of the sample. Since the magnetization of the low-moment tip reverses in the applied magnetic fields, the measured frequency shift contrast was inverted, such that up-magnetized domains consistently appear with a positive (red) frequency shift contrast. The remanent state (Fig.~\ref{fig:fig3}(f)) shows that several spheres remained negatively saturated (see the one highlighted by the black-line circle), while most of the spheres appear to have a \textit{yin-yang} type stripe domain (dashed circle). Given the 3D nature of the curvilinear surfaces investigated here, a stripe domain spirals around the sphere with a decreasing stripe radius, and it culminates with an enlarged ending at the top of the sphere (Fig.~\ref{fig:fig3}(g), $\SI{20}{mT}$). A further increase of the magnetic field leads to a retraction of the domain from the center towards the circumference of the sphere (Fig.~\ref{fig:fig3}(h,i)). These domain structures remain quite robust against further increase in field. Most spheres exhibit a remaining complex magnetic texture at the rim around the sphere despite the large magnetic fields. Magnetic fields as high as $\SI{120}{mT}$ are required to fully saturate the sample in a positive direction
(Fig.~\ref{fig:fig3}(j)). In summary, despite having a curvilinear confined geometry~\cite{sam2025magnetic}, no skyrmions are detected.

\subsection{Local-field induced nucleation of skyrmions}
The tapping mode combined with medium/high-moment tips used in conventional MFM measurements can modify the magnetic states. In particular, it has been shown on multiple occasions that the localized magnetic fields from such MFM tips can lead to nucleation and even motion of skyrmions~\cite{casiraghi2019individual}. Thus, in the following, we apply localized magnetic fields with the tip of an MFM operated in tapping mode under ambient conditions to alter the magnetic textures on the spheres.

\begin{figure*}[]
    \centering
    \includegraphics[width=\linewidth]{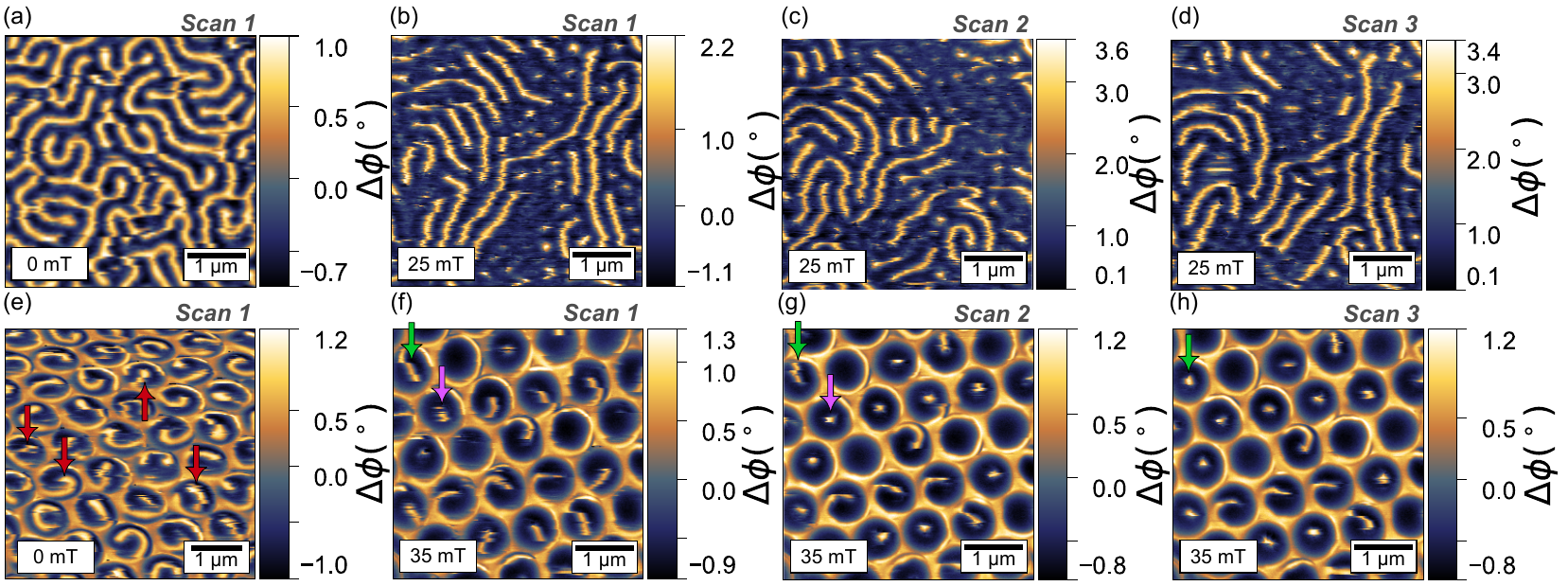}
    \caption{Conventional double-pass MFM under ambient conditions on (a--d) the planar film and (e--h) the spheres. (a) Remanent maze domains of the planar film at $\mu_0H_z = \SI{0}{mT}$. (b--d) Three consecutive scans of the planar film at $\mu_0H_z = \SI{25}{mT}$: the stripe domains are distorted by the tip--sample interaction, and a few metastable skyrmions are nucleated, which disappear again after each scan. (e) Remanent state of the spheres at $\SI{0}{mT}$, showing spiraling stripe domains with sudden relaxation lines and distortions (red arrows). (f--h) Three consecutive scans at $\mu_0H_z = \SI{35}{mT}$: the stripes become unstable (f) and rupture into isolated skyrmions at the top of the spheres upon further scanning (g,h). The purple arrow marks a stripe that collapses into a skyrmion from (f) to (g), whereas the green arrow marks a stripe that remains distorted in (g) and only collapses into a skyrmion in (h). The MFM contrast is given by the phase shift $\Delta\phi$ of the cantilever.}
    \label{fig:fig4}
\end{figure*}

To compare the response of planar and curvilinear films to the local field of the MFM tip, we first imaged the planar film in the same setup. As shown in Fig.~\ref{fig:fig4}(a), the planar film exhibits maze domains at remanence, populated similarly by up and down domains, in good agreement with the STXM results in Fig.~\ref{fig:fig3}(b). When an oop magnetic field of $\SI{25}{mT}$ is applied, the domains evolve into elongated stripes that are distorted by the tip--sample interaction, as shown in Fig.~\ref{fig:fig4}(b). Note that the planar film and the spheres were imaged at oop fields of $\SI{25}{mT}$ and $\SI{35}{mT}$, respectively, for which the best measurement conditions were obtained. Interestingly, a few metastable skyrmions are nucleated by the local field of the tip also in the planar film. However, these skyrmions are not stable and disappear again after each scan, see Fig.~\ref{fig:fig4}(b--d).

The spheres were imaged in the same setup at zero field, as shown in Fig.~\ref{fig:fig4}(e). Here, spiraling (3D) stripe domains are visible. However, a much stronger tip-sample interaction is observed, which can be recognized by the horizontal lines that appear in the phase-shift contrast (red arrows). Furthermore, the stripes appear to jump from one place to another on a sphere. Such tip-induced displacements of magnetic textures are not specific to curved surfaces and are also observed in planar films~\cite{casiraghi2019individual}.

As our high-moment MFM lacked an integrated magnetic field option, we were unable to repeat the stepwise field measurements presented in Fig.~\ref{fig:fig3}. Instead, we used a permanent magnet and non-magnetic spacers to increase the distance between the magnet and the sample to change the magnetic field strength. The magnetic field was measured by an external Hall sensing device as $\SI{35}{mT}$. The actual field experienced by the sample is not known precisely, but we can assume that it is around this value. The first scan of the sample with the applied oop magnetic field is shown in Fig.~\ref{fig:fig4}(f). We also observe a strong tip-to-sample interaction here. However, most of the spiral stripes exhibit horizontal lines and are unstable (purple and green arrows). When we scan for the second consecutive time, as shown in Fig.~\ref{fig:fig4}(g), we now observe that many of the stripes, such as the one marked by the purple arrow, have collapsed into skyrmions stabilized at the top of the spheres. The remaining stripes display fewer relaxation lines, indicating that they are more stable to external perturbations. In contrast, the sphere marked by the green arrow still shows a distorted stripe with relaxation lines. We continue with the scanning, and we find that the skyrmions formed are quite robust, as shown in Fig.~\ref{fig:fig4}(h), and that the stripe marked by the green arrow has now also collapsed into a skyrmion. Overall, the localized magnetic field originating from the coating of the tip, combined with the double-pass scanning method, enables repeated local magnetic field pulses each time the tip passes over a particle, which distort the stripes and then break them into metastable skyrmions. The first scan reveals only $N = 2$ spheres with skyrmions. After the second scan, this number is $N = 10$, and after the third scan this number is $N = 12$ (out of 33 spheres in the field of view). Note that MFM alone cannot determine the chirality and topological charge of these skyrmionic textures. However, since the interfacial DMI of the Pt/Co/Ta multilayer favors homochiral N\'eel domain walls~\cite{wang2019construction}, we expect them to be N\'eel skyrmions, as also obtained in our micromagnetic simulations shown in Fig.~\ref{fig:fig5}(d).

\subsection{Curvature-stabilized spiral states\label{subsec:res_sim}}
\begin{figure*}[]
    \centering
    \includegraphics[width=\linewidth]{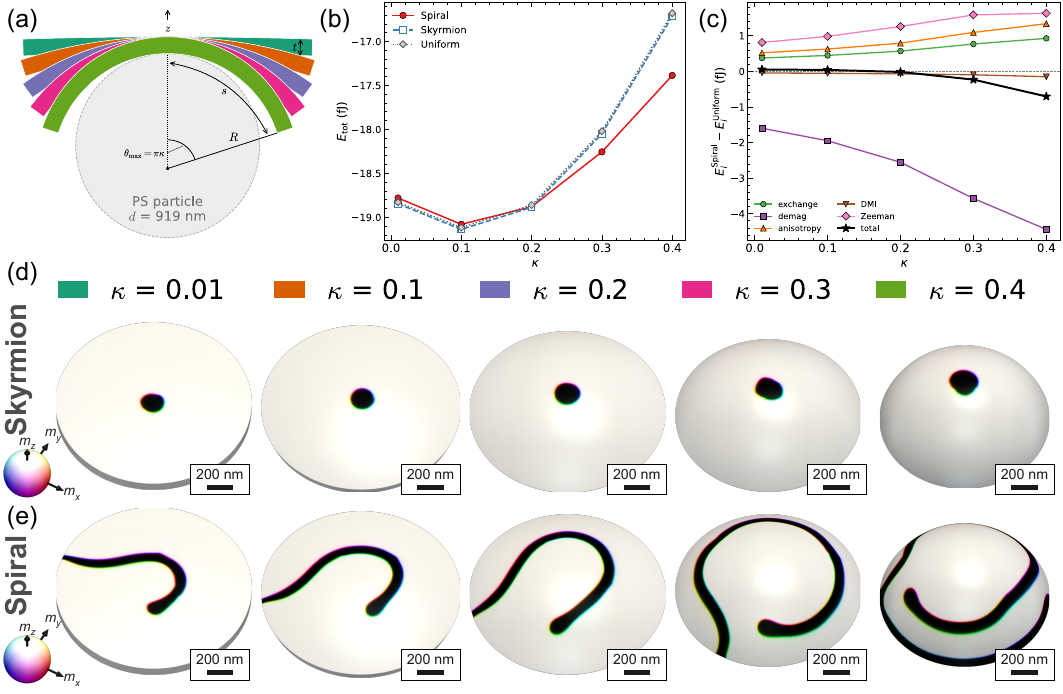}
    \caption{(a) Cross-section of the simulated spherical caps with constant thickness $t$ and geodesic radius $s$ for different curvature parameters $\kappa$, where $R = s/(\pi\kappa)$ is the radius of curvature and $\theta_{\rm max} = \pi\kappa$ the opening half-angle. For comparison, the PS particle with a diameter of $d = \SI{919}{nm}$ is shown in gray. (b) Total energy $E_{\rm tot}$ of the relaxed spiral, skyrmion, and uniform states and (c) energy differences $E_i^{\rm Spiral} - E_i^{\rm Uniform}$ of the individual energy contributions between the spiral and the uniform state as a function of $\kappa$ at $\mu_0 H_z = \SI{120}{mT}$. (d) Relaxed skyrmion and (e) spiral states for increasing $\kappa$. The color code indicates the magnetization direction according to the color sphere: the hue represents the in-plane direction, while white and black indicate a magnetization pointing out of and into the surface, respectively.}
    \label{fig:fig5}
\end{figure*}

To understand why the spirals only decay into skyrmions under the local field of the MFM tip, we compare the energies of the spiral and skyrmion states to that of a uniformly magnetized cap by means of micromagnetic simulations. Figure~\ref{fig:fig5}(a) illustrates the simulated spherical caps, whose curvature is increased from a nearly flat disk ($\kappa = 0.01$) to a strongly curved cap ($\kappa = 0.4$), which most closely resembles our experimental nanocaps, while the film thickness and the geodesic radius are kept constant. For each cap, we relax a spiral, a skyrmion, and a uniform initial state. Since the initial skyrmion expands into labyrinth domains at low fields, we compare the states at $\mu_0H_z = \SI{120}{mT}$. Note that the simulations require much higher fields than the experiment for the stabilization of isolated skyrmions, and that maze domains are obtained at the experimental field values.

The total energies of the relaxed states are presented in Fig.~\ref{fig:fig5}(b). For nearly flat caps with $\kappa \leq 0.2$, the skyrmion is the lowest-energy state, whereas the spiral is even higher in energy than the uniform state. With increasing curvature, the energy of the spiral decreases with respect to both other states. The energy difference $\Delta E = E_{\rm Sk} - E_{\rm Spiral}$ changes sign at $\kappa \approx 0.2$ and reaches $\SI{0.67}{fJ}$ for $\kappa = 0.4$, where the spiral is also $\SI{0.71}{fJ}$ lower in energy than the uniform state. Note that the uniform state is never the lowest-energy state. Thus, for $\kappa \gtrsim 0.2$, the spiral is the preferred state, and the skyrmion is only metastable.

To gain a deeper understanding of this stabilization, we decompose the energy difference between the spiral and the uniform state into its individual contributions in Fig.~\ref{fig:fig5}(c). Compared to the uniform state, the spiral costs Zeeman energy ($\SI{0.81}{fJ}$ for $\kappa = 0.01$ and $\SI{1.63}{fJ}$ for $\kappa = 0.4$), anisotropy energy ($\SI{0.52}{fJ}$ and $\SI{1.34}{fJ}$), and exchange energy ($\SI{0.37}{fJ}$ and $\SI{0.93}{fJ}$). In contrast, the spiral saves demagnetization energy, and this reduction grows much faster with curvature, from $\SI{1.60}{fJ}$ to $\SI{4.44}{fJ}$, such that the spiral becomes energetically favorable for $\kappa \gtrsim 0.2$. Note that the demagnetization energy is computed from the full magnetostatic interaction of the curved film and thus includes the \textit{nonlocal} curvature-induced chiral interactions formulated by Sheka \textit{et al.}~\cite{sheka2020nonlocal}; it is this energy term that is responsible for the stabilization of the spiral state. The interfacial DMI contributes only little compared to the demagnetization energy, and the exchange energy, which contains the local curvature-induced DMI~\cite{gaididei2014curvature, sheka2015curvature}, does not favor the spiral. For the skyrmion, all contributions relative to the uniform state are small and nearly independent of $\kappa$.

The corresponding magnetic states are displayed in Fig.~\ref{fig:fig5}(d,e). As shown in Fig.~\ref{fig:fig5}(d), the skyrmion relaxes for all $\kappa$ into an isolated N\'eel skyrmion at the top of the cap, where the hue of the domain wall rotates once around the core, reflecting its topological charge. In contrast, the spiral relaxes into a single stripe domain that enters the cap at its rim and curls towards the top, see Fig.~\ref{fig:fig5}(e). The stronger the curvature, the more the stripe winds around the top of the cap, resembling the three-dimensional spiral states observed by non-invasive MFM in Fig.~\ref{fig:fig3}. For $\kappa = 0.4$, an additional reversed domain forms along the rim of the cap. Consequently, while the spiral is the lowest-energy state of the strongly curved cap in our simulations, a skyrmion requires an external stimulus to be nucleated, such as the local magnetic field of an MFM tip.

\section{Discussion\label{sec:discussion}}

For the particle sizes used here ($d \approx \SI{919}{nm}$, $R \approx \SI{460}{nm}$), the exchange-driven curvature-induced DMI is expected to be very small. Estimating the exchange-driven $D_{\rm curv} \approx A/R$ with the exchange stiffness $A = \SI{15}{pJ/m}$ used in our simulations yields $D_{\rm curv} \approx \SI{0.03}{mJ/m^2}$, which is more than one order of magnitude smaller than the interfacial DMI of $D = \SI{0.75}{mJ/m^2}$ used in our simulations, which lies within the range of values reported for Pt/Co-based multilayers~\cite{yang2015anatomy, moreau-luchaire2016additive}. Note that this estimate only concerns the geometric, exchange-driven contribution and not a possible curvature-induced modification of the interfacial DMI itself, as proposed in Ref.~\cite{dugato2025curved}; in our asymmetric Pt/Co/Ta stack, a sizable interfacial DMI is present regardless of curvature. Therefore, the skyrmions observed here are unlikely to originate from the local curvature-induced DMI. Instead, our simulations show that the spiral states are stabilized by the magnetostatic interaction on the curved surface, which includes the nonlocal curvature-induced chiral effects~\cite{sheka2020nonlocal}. Thus, the curvilinear geometry gives rise to intrinsic three-dimensional spiral states, as confirmed by our non-invasive in-vacuum MFM measurements, and these spiral states serve as precursors from which skyrmions can be nucleated by the local magnetic field of the MFM tip.

Note that our simulations consider an isolated spherical cap, whereas in the experiment the particles are arranged in a close-packed hexagonal lattice. Since the stabilization of the spiral states is predominantly of magnetostatic origin, the stray fields of the neighboring particles are expected to modify the energy landscape of each cap and might contribute to the sphere-to-sphere variations observed in our MFM measurements, such as the spheres that remain saturated at remanence in Fig.~\ref{fig:fig3}(f). However, already a single cap requires about $8 \times 10^{6}$ tetrahedral elements to resolve the domain walls in the $\SI{84}{nm}$ thick film, such that the simulation of a particle together with its six nearest neighbors is beyond the scope of this work. Nevertheless, we expect that the magnetostatic stabilization of the spiral states found for a single cap remains valid, since the intrinsic spiral states are observed on almost all particles of the lattice.

Our observations are also relevant for recent experimental works that reported skyrmions on curved surfaces~\cite{dugato2025curved, sam2025magnetic}. These studies have provided important insights into how curvature and geometric confinement shape the spin textures on spherical particles, and they relied on MFM as the main imaging technique, complemented by electron holography in Ref.~\cite{dugato2025curved}. Since local tip fields can nucleate skyrmions even in planar films~\cite{casiraghi2019individual, zhang2018magnetic}, our results suggest that a careful assessment of the tip--sample interaction is needed when interpreting such textures, in particular on curved surfaces, where we find metastable skyrmions that are readily nucleated by the tip. In this regard, non-invasive imaging methods provide a valuable complement. The high-sensitivity in-vacuum MFM with low-moment tips employed here reveals only 3D spiral states and no skyrmions. In contrast, the high-moment tip of the conventional MFM nucleates skyrmions both in the planar film and on the spheres. However, under the conditions of our measurements, the skyrmions in the planar film disappear again after each scan, whereas the skyrmions on the spheres remain stable, and their number increases with each consecutive scan. This underscores the benefit of combining non-invasive and conventional imaging techniques to establish the intrinsic magnetic ground state.

Our results indicate that the skyrmions formed upon tip-field application represent metastable states on the curvilinear particles. This is in good qualitative agreement with our simulations, which reveal that the spiral is the lower-energy state on curved caps, while the skyrmion remains metastable. The fact that a local magnetic field is required for the nucleation of skyrmions suggests that a certain energy barrier needs to be overcome, consistent with the metastability of the skyrmions in our simulations. From an experimental perspective, the local field-induced formation of skyrmions opens new possibilities. The MFM tip itself might be used as a writing tool, as demonstrated for artificial spin ice~\cite{gartside2018realization}, where the number of scans is related to the number of skyrmions generated on the curved nanocaps. Alternatively, it might be feasible to generate a local field with a micro- or nanocoil to control skyrmion nucleation and annihilation~\cite{finizio2019deterministic}.

\section{Conclusion\label{sec:conclusion}}
In this work, we investigated $\rm [Pt/Co/Ta]_{12}$ multilayers deposited on self-assembled monolayers of polystyrene particles and compared their magnetic properties to those of planar films. Using non-invasive STXM and high-sensitivity in-vacuum MFM with low-moment tips, we found that the curvilinear films host intrinsic three-dimensional spiral stripe states instead of the maze domains of the planar films, and that no skyrmions were detected in the field range studied. Skyrmions were only observed when the local magnetic field of a conventional high-moment MFM tip ruptured these spirals into single skyrmions located at the top of the particles. While the tip nucleates skyrmions also in the planar film, they disappear again after each scan, whereas the skyrmions on the spheres remain stable and their number increases with each consecutive scan. By means of micromagnetic simulations, we showed that curvature renders a single spiral stripe energetically favorable over an isolated skyrmion, and that this stabilization is predominantly of magnetostatic origin, in agreement with a negligible exchange-driven curvature-induced DMI for the particle sizes studied here. Consequently, the skyrmions on the curved surface are metastable states that require a local stimulus for their nucleation. Our results highlight the value of non-invasive imaging to distinguish intrinsic curvature effects from spin textures induced by the measurement probe. In recent years, experimental studies and numerical modeling of curvilinear and 3D magnetic systems have gained a lot of momentum~\cite{fernandez-pacheco2017threedimensional, ruiz-gomez2025tailoring, gubbiotti20252025}, and the latest developments in focused electron beam-induced deposition and two-photon lithography have revealed new magnetic states and functionalities~\cite{koraltan2025reconfigurable, volkov2024threedimensional, donnelly2022complex}. In this context, self-assembled particle monolayers offer an additional, readily accessible platform to study curvature effects on spin textures and their local manipulation.

\section*{ACKNOWLEDGMENTS}
S.K. and D.S. acknowledge funding from the Austrian Science Fund (FWF) under grant no. I6267 (CHIRALSPIN). S.K. thanks the Vienna Doctoral School in Physics for funding the Mobility Fellowship. S.K. acknowledges support from the Munich Center for Quantum Science and Technology (MCQST), funded by the Deutsche Forschungsgemeinschaft (DFG, German Research Foundation) under Germany's Excellence Strategy -- EXC-2111/2. S.K., T.G., J.J., A.F.P. acknowledge funding from the European Research Council (ERC) under the European Union’s Horizon 2020 research and innovation programme, grant agreement no. 101001290 (3DNANOMAG). M.A. gratefully acknowledges funding from Deutsche Forschungsgemeinschaft (DFG, German Research Foundation) grant no. 507821284. J.S. acknowledges financial support and a scholarship received from the Bavarian Research Foundation (DOK-193-22). The work performed at Empa was supported through the SNF Lead Agency project 200021E 211828 “Binary information represented by chiral spin textures”. We thank the Helmholtz-Zentrum Berlin für Materialien und Energie for the allocation of synchrotron radiation beamtime. R.K. and C.A. acknowledge that this research was funded in whole or in part by the Austrian Science Fund (FWF) [10.55776/P34671 and 10.55776/PIN1629824].

S.K., D.S. and M.A. conceived the project, S.K. and J.S. sputtered the magnetic layers and performed the ambient MFM measurements, M.K. performed the nanoparticle lithography, E.D., D.R., R.P.P., H.J.H. improved the vacuum MFM setup, and performed the measurements with S.K. and J.S., S.K., T.G. performed the VSM with support from J.J. and A.F.P., S.K., T.G., J.J., M.W., and S.W. performed the STXM measurements, R.K. and C.A. wrote and improved the micromagnetic code and simulations, S.K. and M.A. wrote the initial draft of the manuscript. All authors reviewed and contributed to the final version of the manuscript.  

\section*{Data availability statement}
The data that support the findings of this article are openly available at https://doi.org/10.5281/zenodo.17845870.

\bibliography{references}

\end{document}